\documentclass[prl,aps,twocolumn,superscriptaddress,showpacs]{revtex4}
\usepackage[T1]{fontenc}
\usepackage{ae}   
\pdfoutput=1
\usepackage{wrapfig}
\usepackage{float}
\usepackage{amsmath}
\usepackage{amssymb}
\usepackage{amsfonts}
\usepackage[pdftex]{graphicx}           
\usepackage{array}      
\usepackage{bm}         
\usepackage{dsfont}     
\usepackage{ulem}
\usepackage{color}
\usepackage{hyperref}
\usepackage{units}
\usepackage{gensymb}

\renewcommand{\vec}[1]{\textbf{\textit{#1}}}

\begin{document}
\title{Scanning Tunneling Spectroscopy on SrFe$\mathrm{_2}$(As$\mathrm{_{1-x}}$P$\mathrm{_x}$)$\mathrm{_2}$}
\author{Jasmin Jandke}
\affiliation{Physikalisches Institut, Karlsruher Institut f\"ur Technologie, 76131
Karlsruhe, Germany}
\author{Petra Wild}
\affiliation{Physikalisches Institut, Karlsruher Institut f\"ur Technologie, 76131
Karlsruhe, Germany}
\author{Michael Schackert}
\affiliation{Physikalisches Institut, Karlsruher Institut f\"ur Technologie, 76131
Karlsruhe, Germany}
\author{Shigemasa Suga}
\affiliation{Institute of Scientific and Industrial Research, Osaka University, 8-1 Mihogaoka, Ibaraki, Osaka 567-0047, Japan}
\author{Tatsuya Kobayashi}
\affiliation{Department of Phyiscs, Graduate School of Sciences, Osaka University, Osaka 560-0043, Japan}
\author{Shigeki Miyasaka}
\affiliation{Department of Phyiscs, Graduate School of Sciences, Osaka University, Osaka 560-0043, Japan}
\author{Setsuko Tajima}
\affiliation{Department of Phyiscs, Graduate School of Sciences, Osaka University, Osaka 560-0043, Japan}

\author{Wulf Wulfhekel}
\affiliation{Physikalisches Institut, Karlsruher Institut f\"ur Technologie, 76131
Karlsruhe, Germany}
\date{\today }
\begin{abstract}
We investigated SrFe$\mathrm{_2}$(As$\mathrm{_{1-x}}$P$\mathrm{_x}$)$\mathrm{_2}$ single crystals with four different phosphorus concentrations x in the superconducting phase (x = 0.35, 0.46) and in the magnetic phase (x = 0, 0.2). The superconducting samples display a V-shaped superconducting gap, which suggests nodal superconductivity. Furthermore we determined the superconducting coherence length by measuring the spatially resolved superconducting density of states (DOS). Using inelastic tunneling spectroscopy we investigated excitations in the samples with four different phosphorus concentrations. Inelastic peaks are related to bosonic modes. Phonon and non-phonon mechanism for the origin of these peaks are discussed.
\end{abstract}
\maketitle
\section{Introduction}
After the discovery of the first iron-based superconductors with LaOFeAs in 2008 \cite{Kamihara2008} a large amount of new iron-based superconductors has been found. In spite of intensive investigations using a variety of techniques the superconducting pairing mechanism and the microscopic origin of the magnetism in iron-based superconductors is still under debate \cite{dai2015magnetic}.
Nevertheless, combining different results accumulated so far, the knowledge about the physical properties of the iron pnictides has progressed. Key properties in view of the determination of the pairing mechanism are the pairing symmetry and the investigation of bosonic excitations which probably act as the 'pairing-glue' for the Cooper-pair formation. The characteristic length scale of the Cooper pairs is called the superconducting coherence length.

In contrast to cuprates and conventional superconductors, the superconducting gap distribution in iron based superconductors is rather diversified \cite{zhang2015electron}. Nodeless isotropic gap distributions have been observed in Ba$\mathrm{_{1-x}}$K$\mathrm{_x}$Fe$\mathrm{_2}$As$\mathrm{_2}$, Ba$\mathrm{_2}$Fe$\mathrm{_{2-x}}$Co$\mathrm{_x}$As$\mathrm{_2}$, LiFeAs, NaFe$\mathrm{_{1-x}}$Co$\mathrm{_x}$As$\mathrm{_2}$, FeTe$\mathrm{_{1-x}}$Se$\mathrm{_x}$ \cite{zhang2015electron,Ding2008,Terashima2009,Miao2012,Liu2011,Umezawa2012,Borisenko2012} whereas strong signatures of nodal superconducting gap have been reported by various experimental techniques in LaOFeP \cite{Fletcher2009}, LiFeP \cite{Hashimoto2012}, underdoped Ba$\mathrm{_{1-x}}$K$\mathrm{_x}$Fe$\mathrm{_2}$As$\mathrm{_2}$ \cite{Zhang2012,Reid2011}, BaFe$\mathrm{_{2-x}}$Ru$\mathrm{_x}$As$\mathrm{_2}$ \cite{Qiu2012}, KFe$\mathrm{_2}$As$\mathrm{_2}$ \cite{Dong2010}, FeSe \cite{Song2011} and BaFe$\mathrm{_2}$(As$\mathrm{_{1-x}}$P$\mathrm{_x})_2$ \cite{Zhang2012,Hashimoto2010,Qiu2012,zhang2015electron,Nakai2010,Kim2010,Wang2011,Yamashita2011}. Similar nodal gap structures were expected for other nodal compounds like the phosphorus-based iron pnictides \cite{zhang2015electron}. Indeed, for the optimally doped compound SrFe$\mathrm{_2}$(As$\mathrm{_{0.65}}$P$\mathrm{_{0.35}}$)$\mathrm{_2}$ evidence suggested nodal superconductivity by doing Phosphorus-31 nuclear magnetic resonance ($^{31}$P-NMR), specific heat and London penetration depth measurements \cite{Dulguun2012,Murphy2013, Takahashi2012}. Within this paper, we present our results of scanning tunneling microscopy/spectroscopy (STM/STS) measurements on  isovalently doped SrFe$\mathrm{_2}$(As$\mathrm{_{1-x}}$P$\mathrm{_x}$)$\mathrm{_2}$ confirming nodal superconductivity for the optimally doped (x=0.35) and overdoped (x=0.46) compound with a hint for $s_{\pm }$ pairing symmetry. While the undoped compounds have been investigated with STM \cite{Dreyer2011,Dutta2015}, no STM/STS were performed so far on doped samples suffering from a possible doping inhomogeneity. Here, STM provides a useful tool for investigation of the surface morphology and the superconducting gap by means of elastic tunneling. Inelastic tunneling is a precise tool to reveal the underlying bosonic structure in conventional superconductors \cite{Schackert2015, Jandke2015}. We, here, apply inelastic tunneling to unconventional superconductors. Since for the optimally doped compound - SrFe$\mathrm{_2}$(As$\mathrm{_{0.65}}$P$\mathrm{_{0.35}}$)$\mathrm{_2}$ - the superconducting transition temperature $T_c$ is about \unit[30]{K} \cite{Kobayashi2012}, it is unlikely that phonons are the particles responsibe for the Cooper pairing \cite{McMillan1968,Stojchevska2010}. Hence, the investigation of other excitations which could act as 'pairing-glue' is important.  Especially, for the investigation of doped samples STM is an appropriate method to address these samples due to its ability to spatially resolve the DOS. This allows us to directly determine the coherence length in the superconducting compounds from measurements of the local DOS as well.

Depending on the phosphorus concentration and temperature, SrFe$\mathrm{_2}$(As$\mathrm{_{1-x}}$P$\mathrm{_x}$)$\mathrm{_2}$ can either be in the magnetic phase, in the superconducting phase or in the normal conducting phase \cite{Kobayashi2012}. In the present paper we investigated SrFe$\mathrm{_2}$(As$\mathrm{_{1-x}}$P$\mathrm{_x}$)$\mathrm{_2}$ by performing STS for four different doping concentrations. In the first part of the paper we show the results of the optimally doped (x=0.35) and overdoped (x=0.46) superconducting compound. The method for measuring  the superconducting gap and the coherence length will be explained therein. In the second part, possible inelastic excitations of these compounds are compared to those of the  magnetic compound (x=0.2) and the parent compound (x=0). 

\section{Methods and Results}
\begin{figure}[H]
\includegraphics[scale=0.4]{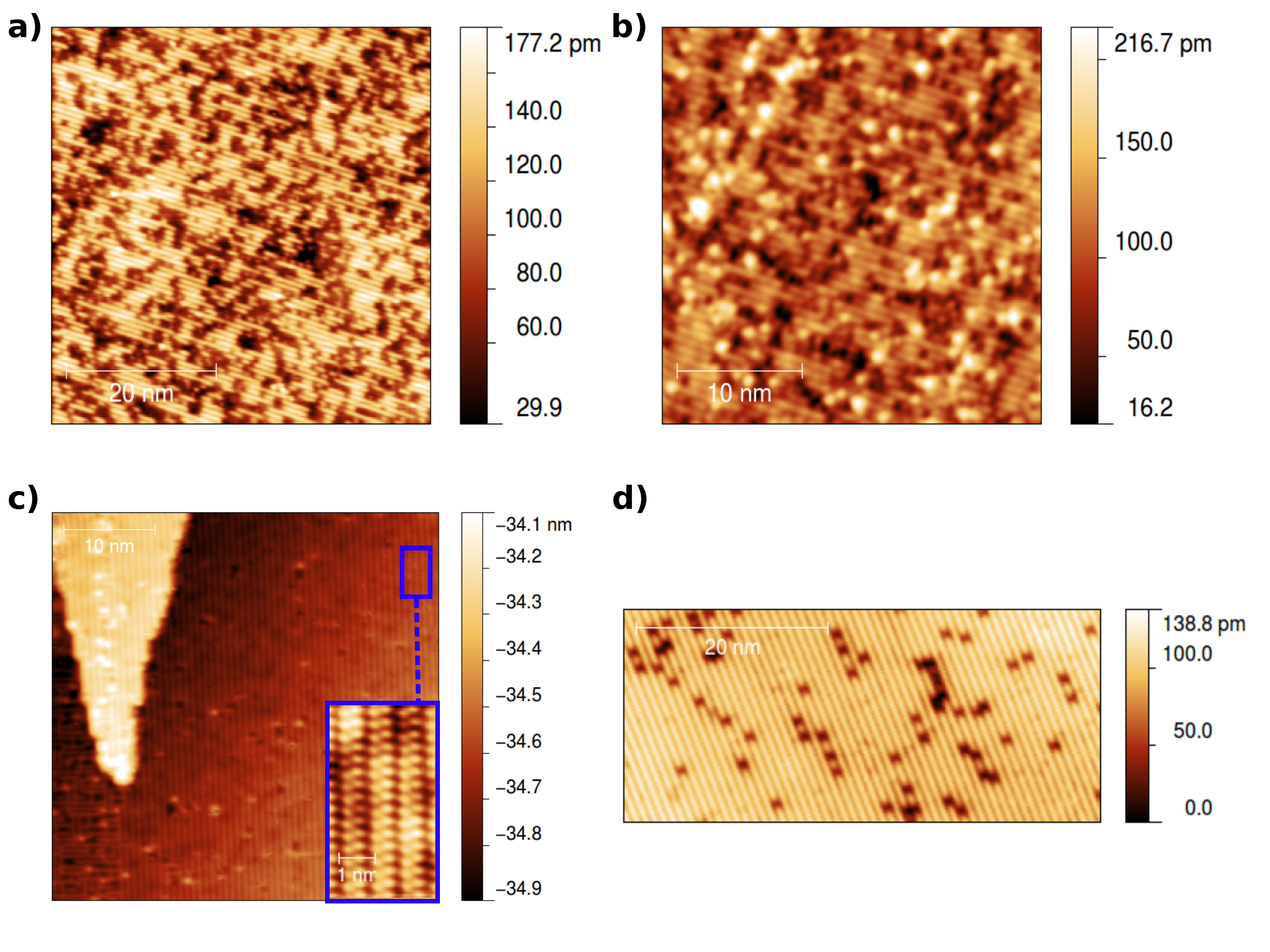}
\caption{(Color online) Topographic images for a) the optimally doped compound (x=0.35), (U=\unit[1]{V}, I=\unit[100]{pA}), b) the overdoped compound (x=0.46), (U=\unit[7]{mV}, I=\unit[1]{nA}), c) the underdoped compound (x=0.2), (U=\unit[80]{mV}, I=\unit[1]{nA} and U=\unit[6.4]{mV}, I=\unit[2]{nA} for inset), d) the parent compound with x=0 (U=\unit[600]{mV}, I=\unit[1nA]).}
\label{fig1}
\end{figure}

\begin{figure}[H]
\includegraphics[scale=0.2]{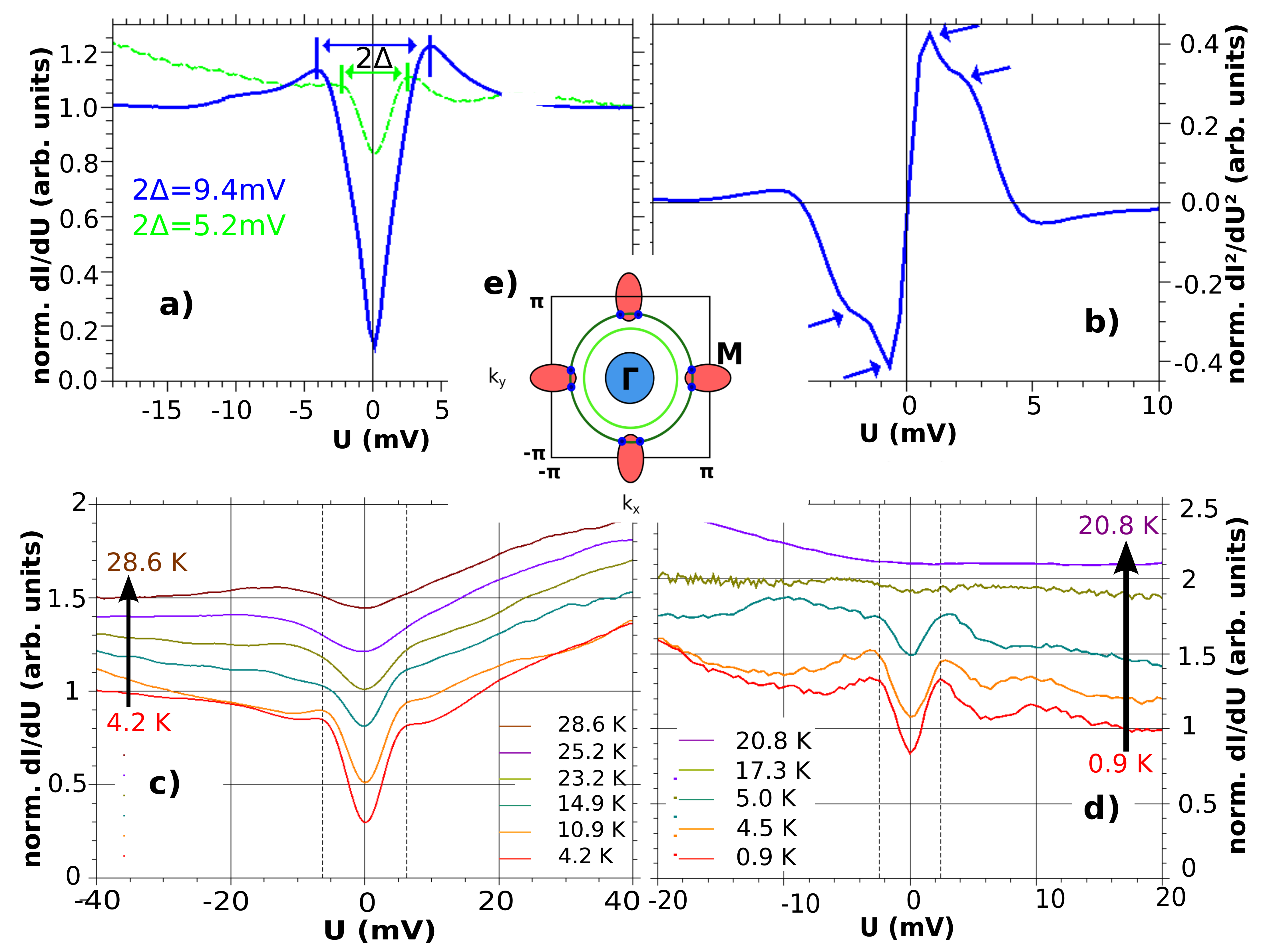}
\caption{(Color online) a)  Spatially averaged superconducting gap over many spectra for the optimally doped (blue) and overdoped (dashed light-green) compound at $\mathrm{T}=\unit[0.8]{K}$ ( $\mathrm{I_{set}}$=\unit[2.15]{nA}, $\mathrm{U_{mod}}$=\unit[1]{mV}), b) numerical derivative of the optimally doped spectrum (blue curve in a)), c) Temperature dependence of the superconducting gap for the optimally doped compound (x=0.35) ($\mathrm{I_{set}}$=\unit[5]{nA}), d) Temperature dependence of the superconducting gap for the overdoped compound (x=0.46) ($\mathrm{I_{set}}$=\unit[3.5]{nA}), e) Nodal line of the order parameter in the Brillouin zone: for the green line $\Delta _1/ \Delta _2 =1$ whereas for the dark green line $\Delta _1/\Delta _2 <1$ and the nodal line intercepts the electron-like Fermi surfaces. Blue dots mark the positions where the gap-size on the Fermi surface is zero. Adapted from \protect\cite{Hoffman2011}}
\label{fig2}
\end{figure}

The SrFe$\mathrm{_2}$(As$\mathrm{_{1-x}}$P$\mathrm{_x}$)$\mathrm{_2}$ single crystals were synthesized using the self flux method \cite{Kobayashi2012}. All investigated crystals were cleaved at $p \sim 1\cdot \unit[10^{-10}]{mbar}$ at $\unit[77]{K}$ and afterwards immediately transferred to the STM-chamber. Cleavage posts were glued onto the samples using a triple-axis manipulator \footnote{3D micrometer-drive lift from VIC International, Tokyo, Japan.}.
Measurements were done with a home-build Joule Thomson low temperature STM (JT-STM) \cite{Zhang2011} at about $\unit[0.8]{K}$. The JT-STM contains a magnet which allows to enter the Shubnikov phase of SrFe$\mathrm{_2}$(As$\mathrm{_{1-x}}$P$\mathrm{_x}$)$\mathrm{_2}$. Topographic images for the four different compounds are shown in Fig.\ref{fig1}. Due to covalent bonds between Fe and As atoms the FeAs-layers are assumed to remain intact during the cleavage. In fact the cleavage occurs either between the As and Sr layer or within the Sr layer leaving half of the atoms of the Sr-layer on the topmost layer forming a $(\sqrt{2}\times \sqrt{2})$ or $(2\times 1)$-reconstruction \cite{Hoffman2011,Dreyer2011}. 
In our measurements, a $(2\times 1)$-reconstruction was present for all samples. However, the samples with higher doping concentrations (x=0.35, 0.46) showed more defects and impurities coming from the higher phosphorus concentration. The tunneling spectra (cf Fig. \ref{fig2}a)) are spatially averaged over many spectra. The width of the superconducting gap $2\Delta $ was determined by the positions of the quasiparticle peaks which occur at $\Delta =\pm \unit[4.7]{mV}$ for the optimally doped compound and at $\Delta =\pm \unit[2.6]{mV}$ for the overdoped compound as indicated in the figure. Even though they were measured at T=\unit[0.8]{K} the superconducting gap is V-shaped and does not go completely down to zero at zero bias.  Nevertheless, the temperature dependence of the gap (see Fig. \ref{fig2} c)/d)) and the appearance of a vortex lattice by applying a magnetic field (see Fig. \ref{wulfhekelf1_f3} a)/c)) proof that the gap is indeed due to superconductivity.
In Fig.\ref{fig2} a) the superconducting gap of the optimally doped compound is compared to that of the overdoped compound.

Iron based superconductors, as well as the present system, obey a complicated Fermi surface characterized by five sets of Fermi sheets arising from the d-orbital of iron. Two of them form electron like pockets at the M point, while the other three build up hole like pockets centered at the $\Gamma $ point \cite{Wen2011} (cf. Fig. \ref{fig2}e)). The different Fermi surfaces give rise to different gap values, characteristic for multi-band superconductors. Indeed, our optimally doped sample shows a double-gap like feature. This can be seen in Fig.\ref{fig2}b), where we observe not only one dip-peak pair but also two features inside the V-shaped gap. The dip and peak features closest to zero bias correspond to the largest slope of the superconducting DOS shown in blue in Fig.\ref{fig2}a). The peaks and dips, marked by arrows, correspond to local maxima of the slope of the superconducting DOS (blue line in Fig.\ref{fig2}a)). Further hints for multiple gap values have  been obtained experimentally in angular resolved photoemission spectroscopy (ARPES) studies on Ba122-K40 \cite{Shimojima2011} as well as in STM experiments on Ba122-Co6 \cite{Teague2011}, \cite{Hoffman2011}. 

In Fig. \ref{fig2}e) a simpler sketch of the Fermi surface is illustrated only depicting one kind of electron and hole pockets, respectively. The electron and hole-like Fermi surfaces are linked via scattering processes involving antiferromagnetic spin fluctuations \cite{Mazin2008}. In conjunction of this scattering process, if it was the mechanism of Cooper pair formation, an unconventional $s_{\pm }$-symmetry of the order parameter was proposed \cite{Mazin2008} which exhibits different sign on the electron and hole pockets. In this extended model the order parameter $\Delta$ can be described by two terms \cite{Song2011}
\begin{equation}
\Delta _{s_{\pm }}(\vec{k})=\Delta _1\:\mathrm{cos}\;\!{k_x}\:\mathrm{cos}\;\!{k_y}+\Delta _2\:(\mathrm{cos}\;\!{k_x}+\mathrm{cos}\;\!{k_y}),
\end{equation}
where $\vec{k}$ is a vector in reciprocal space. The ratio of the parameters $\Delta _1$ and $\Delta_2$ specifies which of them dominates resulting in a nodeless or nodal order parameter. The order parameter changes its sign on an almost circular nodal line centered around the $\Gamma $ point (see Fig. \ref{fig2}e)). Depending on the ratio $\Delta _1/\Delta _2$, the nodal line of the full function $\Delta _{s_\pm }(\vec{k})$ may or may not pass through the electron-like Fermi surfaces \cite{Hoffman2011}. If it passes the Fermi surface a nodal order parameter occurs. From the experimental side the $s^{\pm }$-symmetry of the order parameter is supported by STM experiments exploiting QPI in connection with an additional magnetic field in Fe(Se,Te) \cite{Hanaguri2010}.

If the order parameter vanishes at some points on the Fermi surface, the quasiparticle density of states, which is averaged over all momenta, is thus not fully gapped anymore and a V-shape gap emerges. The results shown in Fig. \ref{fig2} prove, that the system SrFe$\mathrm{_2}$(As$\mathrm{_{1-x}}$P$\mathrm{_x}$)$\mathrm{_2}$ possesses a nodal superconducting gap, suggesting $s_{\pm }$-symmetry in this system. While the d-symmetry always induces a nodal gap due to symmetry in cuprates, in the pnictides the appearance of a V-shaped gap depends on the details of the compound under investigation. This effect was already observed experimentally in the comparison of tunneling spectra of FeTe$_\mathrm{1-x}$Se$_x$ which are fully gapped \cite{Hanaguri2010} and FeSe, where a V-shaped gap was observed \cite{Song2011}.

Next, we determined the coherence length of the optimally and overdoped compounds by using two different methods for both of them. On one hand we applied the power spectral density function (PSDF) on a measured superconducting gap map and on the other hand we applied a magnetic field to extract the coherence length from a vortex. First we describe the PSDF-method.

\textit{PSDF-method}.

\begin{figure}[H]
\begin{center}
\includegraphics[width=0.5\textwidth]{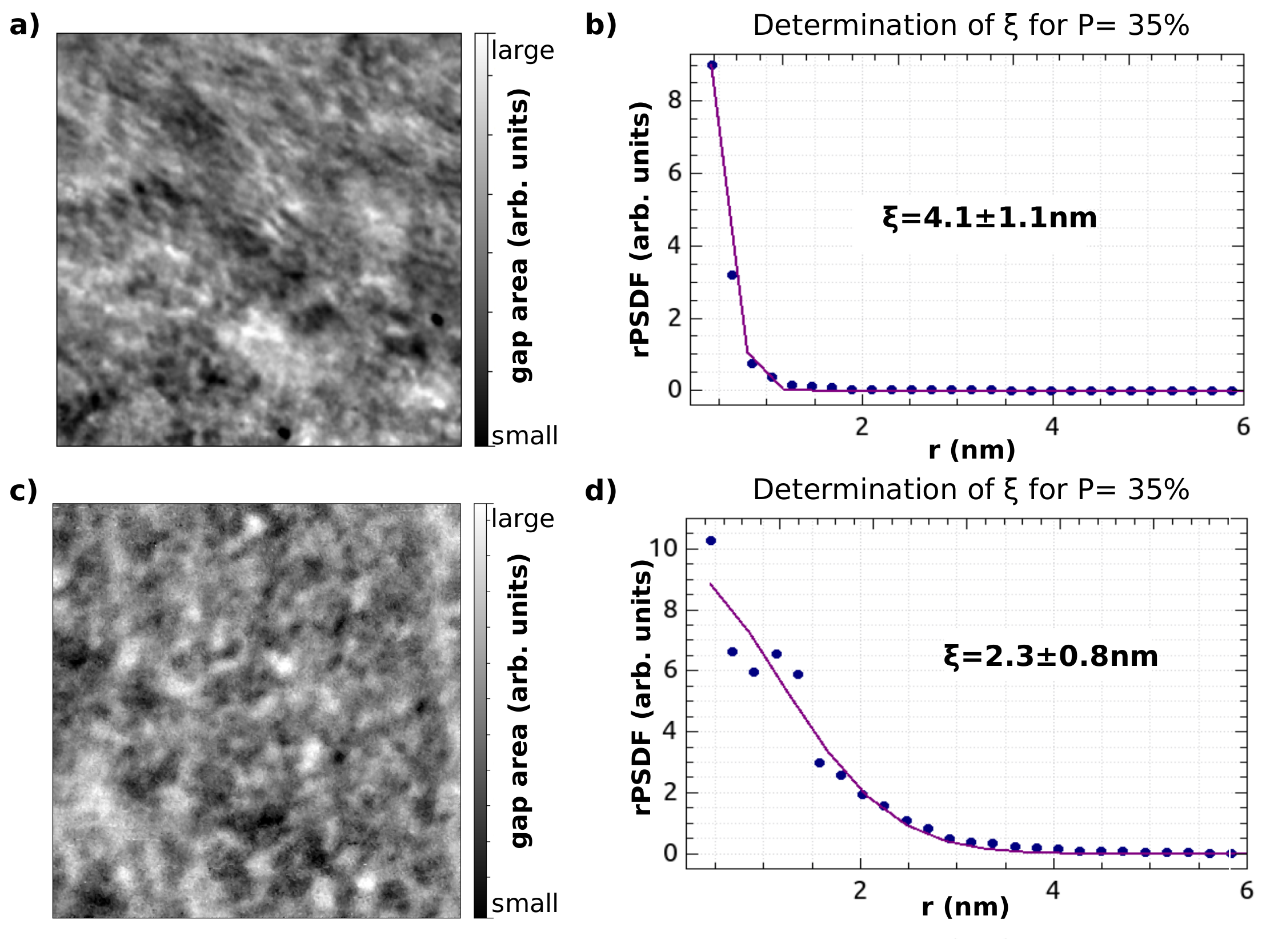}
\end{center}
\caption{(Color online) a) Spatially resolved map of the superconducting gap area for the optimally doped compound (x=0.35) ($\unit[30\times 30]{nm^2}, \unit[256\times 256]{pixel}$), b) calculated radial resolved PSDF of the gap map shown in a) (dots) and the applied fit shown as solid line,  c) spatially resolved map of the superconducting gap area for the overdoped compound (x=0.46) ($\unit[35\times 35]{nm^2}, \unit[256\times 256]{pixel}$), d) calculated radial resolved PSDF of the gap map shown in c) (dots) and the applied fit shown as solid line}
\label{wulfhekel1_f2}
\end{figure}

\begin{figure}
\begin{center}
\includegraphics[width=0.5\textwidth]{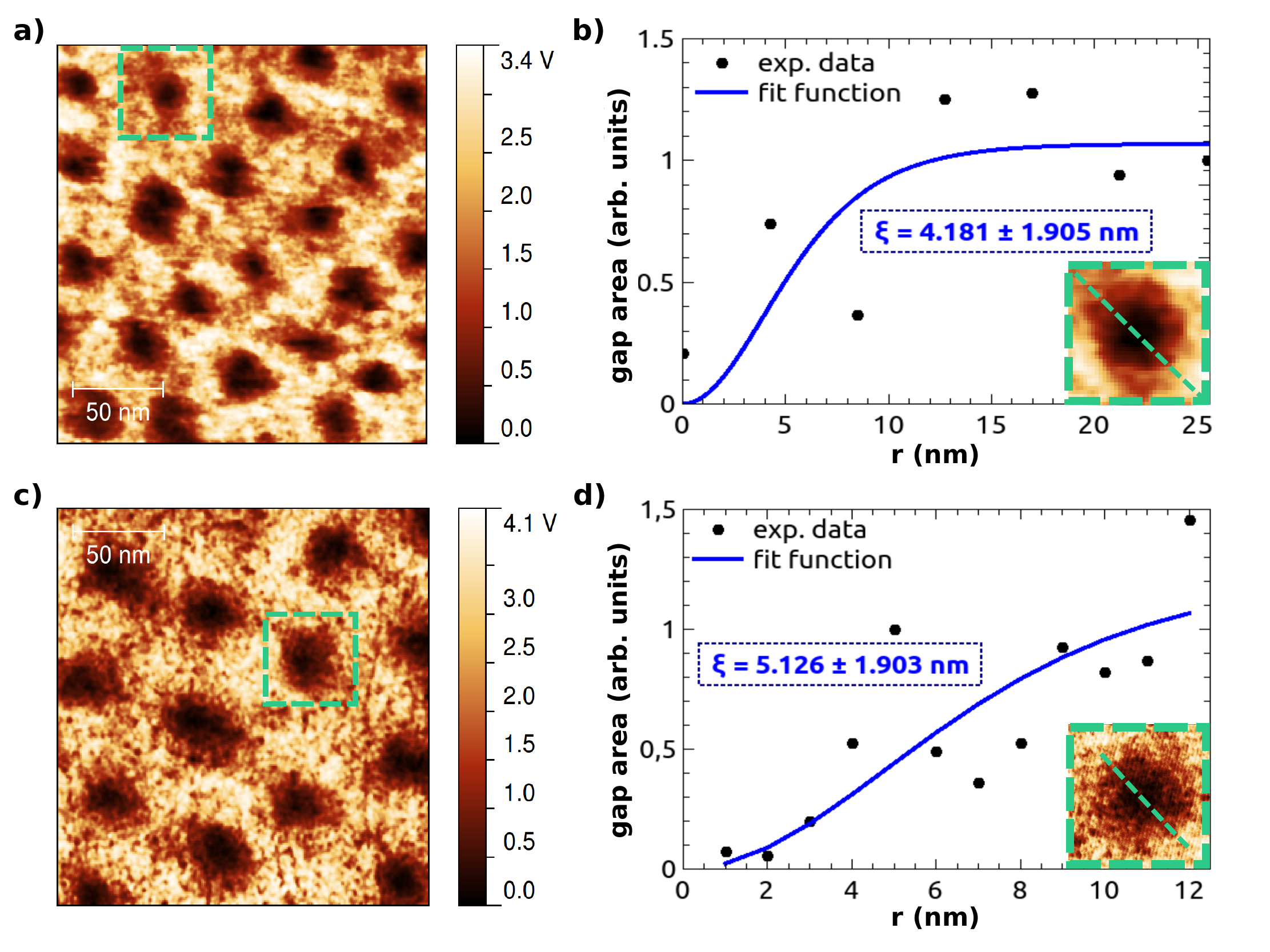}
\end{center}
\caption{(Color online) a) Vortex lattice of the optimally doped compound recorded as $d^2I/dU^2$-map at $U=\unit[2]{mV}$, b) calculated superconducting gap area along a line through a vortex which is marked with a green rectangle in a) and b) (dots), applied fit function (solid line), c) Vortex lattice of the overdoped compound recorded as $d^2I/dU^2$-map at $U=\unit[1.2]{mV}$, d) calculated superconducting gap area along a line through a vortex which marked with a green rectangle in c) and d) (dots), applied fit function (solid line)}
\label{wulfhekelf1_f3}
\end{figure}

Due to the random phosphorus doping, inhomogeneities within the sample lead to spatial variations of the superconducting gap on the minimal length scale set by the coherence length. Thinking about conventional superconductors, the superconducting ground state occurs due to a large number of overlapping Cooper-pair wavefunctions where the phase of the Cooper-pair is the same as the one of the superconducting ground state. For SrFe$\mathrm{_2}$(As$\mathrm{_{1-x}}$P$\mathrm{_x}$)$\mathrm{_2}$ we assume that the spatial variation of doping concentration is convoluted with the wavefunction of the Cooper-pairs of the extension of the coherence length $\xi $. The wavefunction of a Cooper-pair can be described by a Gaussian function $g(x,y)=\frac{1}{2\pi \sigma ^2}e^{-\frac{x^2+y^2}{2\sigma ^2}}$. In order to extract $\xi =2\sigma \sqrt{2\mathrm{ln}(2)}$  we did spatially resolved STS measurements over an area of $\unit[30]{nm}\times \unit[30]{nm}^2$ with $256\times 256$ spectra. For each spectrum the gap area was evaluated using the trapezoidal rule. Finally, a map of the superconducting gap area was generated (see Fig.\ref{wulfhekel1_f2}a),\ref{wulfhekel1_f2}c)).
Such a map shows bright and dark areas. Bright areas correspond to large values for the superconducting gap area and hence to a pronounced  superconducting behaviour. In contrast, in the darker areas the superconductivity is suppressed. On these images we applied the radial resolved PSDF, where the PSDF is the square of the absolute value of the Fourier transformation of a function (PSDF$=\left| \mathcal{F}(f(x,y))\right| ^2$)\cite{Lee1995,Hertel2007}. We can calculate the coherence length assuming such an image consists of randomly distributed superconducting areas convoluted with a Gaussian distribution g(x,y)
\begin{equation}
\left| \mathcal{F}(image)\right| ^2= \underbrace{\left| \mathcal{F}(random)\right| ^2}_{const}* \left| \mathcal{F}(g(x,y))\right|  ^2.
\end{equation}
Using the relationship $\xi =2 \sigma \sqrt{2ln(2)}$ the coherence length can be extracted from $\mathcal{F}(g(x,y))$ as shown in Fig.\ref{wulfhekel1_f2}b) and \ref{wulfhekel1_f2}d). By averaging several measurements in different regions on the surface for the optimally doped compound as well as for the overdoped compound we obtain for the in-plane superconducting coherence length a value of $\xi _{x=0.35}= \unit[4.1\pm 1.1]{nm}$ for the optimally doped compound and $\xi _{x=0.46}=\unit[2.3\pm 0.8]{nm}$ for the overdoped compound.
In order to verify these results we additionally determined the coherence length by using a second method which we call the Vortex-method.

\textit{Vortex-method}

By applying a magnetic field of 1T, the optimally doped as well as the overdoped compound enter the Shubnikov phase. The vortex-lattice could be resolved in $d^2I/dU^2 $-maps. For this, we set the  bias voltage to \unit[2]{mV} and \unit[1.2]{mV} respectively for the optimally doped and overdoped compound. These bias voltages correspond to the largest slope of the superconducting DOS in the optimally/overdoped compound, visible as peaks in the second derivative of the tunneling current. Recording $d^2I/dU^2 $-maps at this bias voltages allows us to distinguish superconducting and normal conducting areas. The superconducting areas appear brighter due to the pronounced gap around the Fermi energy. The vortex-lattice is shown in Fig.\ref{wulfhekelf1_f3}a) and Fig.\ref{wulfhekelf1_f3}c). 
In the vortex core, the material is in the normal state, where the superconducting order parameter is completely suppressed. With increasing distance $r$ from the vortex core the superconducting order parameter $\Delta$ rises according to $\Delta (r) = \mathrm{tanh}(\frac{r}{\sqrt{2}\xi })$ and finally converges to the value in the superconducting state \cite{Poole1995}. By performing STS on a line through a vortex it is possible to extract the coherence length by fitting the measured data with a function $f(r)=a\cdot \mathrm{tanh}^2(\frac{r}{\sqrt{2}\xi})+c$  since the DOS is related to $\left| \Delta (r)\right| ^2$. In order to be consistent with the PSDF-method, we again evaluated the superconducting gap area, but now, within the vortex method, we did the procedure for each point along a line through the vortex (see lower right of Fig.\ref{wulfhekelf1_f3} b) and d)). The obtained data are plotted against the distance from the vortex as shown in Fig.\ref{wulfhekelf1_f3}b) and Fig.\ref{wulfhekelf1_f3}d). As a result we obtain $\xi=\unit[4.181\pm 1.905]{nm}/\xi=\unit[5.126\pm 1.903]{nm}$ for the optimally/overdoped compound. The agreement to the PSDF-method is rather good in the case of the optimally doped compound. Nevertheless, the PSDF-method is more accurate as it contains information of a higher number of local spectra. The vortex method requires to measure along a line through the vortex, which should not move during the measurement. Especially for the optimally doped compound the vortices were mobile even during scanning and were not well-pinned as can also be seen in the vortex lattice in Fig.\ref{wulfhekelf1_f3}a). 

Finally, a theoretical estimation for the coherence length can be made by using the well established relation $H_{c2}=\frac{\Delta _0}{2\pi \xi ^2}$ \cite{Buckel2004}. For the optimally doped compound the upper critical field $H_{c2}$ is about \unit[60]{T} \cite{Takahashi2012}. This results in $\xi _{theo}\approx \unit[2.34]{nm}$, which fits well with our results. Furthermore, if we compare them to values for the coherence length of similar systems they are in the same order of magnitude \cite{Beek2010}.

\begin{figure}[H]
\includegraphics[width=0.4\textwidth]{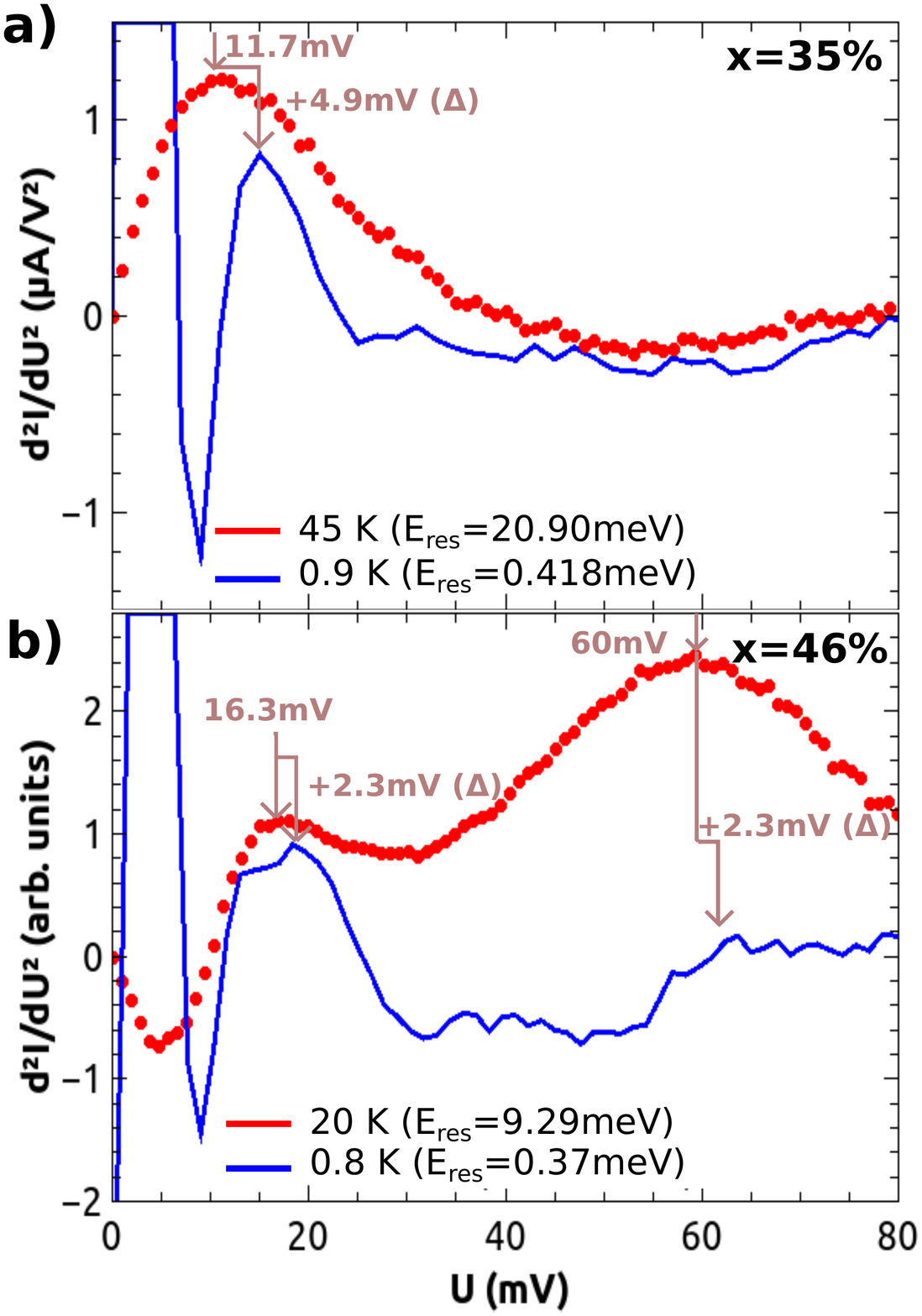}
\caption{(Color online) a) Optimally doped compound:  $d^2I/dU^2$-spectrum for $T>T_c$ (red dots) and for $T<T_c$ (blue line) ($\mathrm{U_m}=\unit[4.3]{mV}$, $\mathrm{I_{set}}=\unit[21]{nA}$), b) Overdoped compound: $d^2I/dU^2$-spectrum for $T>T_c$ (red dots) and for $T<T_c$ (blue line)}
\label{inel1}
\end{figure}

In a last step, we performed inelastic tunneling spectroscopy (ITS) for the four different doping concentrations. For the superconducting compounds (x=0.35/0.46) ITS spectra were measured in the normal conducting state ($T>T_c$) and the superconducting state ($T<T_c$) respectively. In Fig.\ref{inel1} a) spectra for the optimally doped compound are shown. In the normal conducting state (red line) a peak at \unit[11.7]{mV} is visible which could be assigned to an optical phonon mainly arising from the atomic displacements of As and Fe atoms \cite{Litvinchuk2008,Kobayashi2011}. Furthermore, this peak exists in the superconducting state as well (blue line) where it is shifted about \unit[4.9]{mV} to higher energies due to the existence of the sc-gap of $\Delta $=\unit[4.7]{mV}. The results of the same measurement for the overdoped compound are shown in Fig. \ref{inel1}b). In this case a peak around \unit[16.3]{mV} is visible in the normal state (red line) which could either be refered to the same phonon as in Fig.\ref{inel1} a) or to another optical phonon in this system related to atomic displacements of the Sr atoms \cite{Litvinchuk2008}. For the superconducting state this peak is shifted about \unit[2.3]{mV} to higher voltages, again due to the existing  superconducting gap, which is \unit[2.6]{mV} in the case of the overdoped compound. Furthermore, for the overdoped compound (see Fig.\ref{inel1} b)), an additional peak is visible at \unit[60]{mV} for the superconducting as well as for the normal state. Since the Van-Hove singularities in the phonon dispersion relation occur only in an energy range of $\unit[13-40]{meV}$ for the parent compound,  \cite{Litvinchuk2008, Zbiri2010}, this peak cannot be assigned to any phonon. A possible explanation for this peak would be a magnon within this energy. Using optical techniques, excitations at $\unit[68]{meV}$ have been found in the parent compound and have been assigned to magnons \cite{Hancock2010}. The difference of \unit[8]{meV} compared to our measurements could be explained due to the phosporous concentration in our overdoped compound or the energy resolution of only about $\unit[9]{meV}$ for the spectra above $T_c$. Hence, we suggest that the observed dip-peak at $\unit[60]{meV}$ in the overdoped compound is due to a magnon.

\begin{figure}[H]
\includegraphics[width=0.5\textwidth]{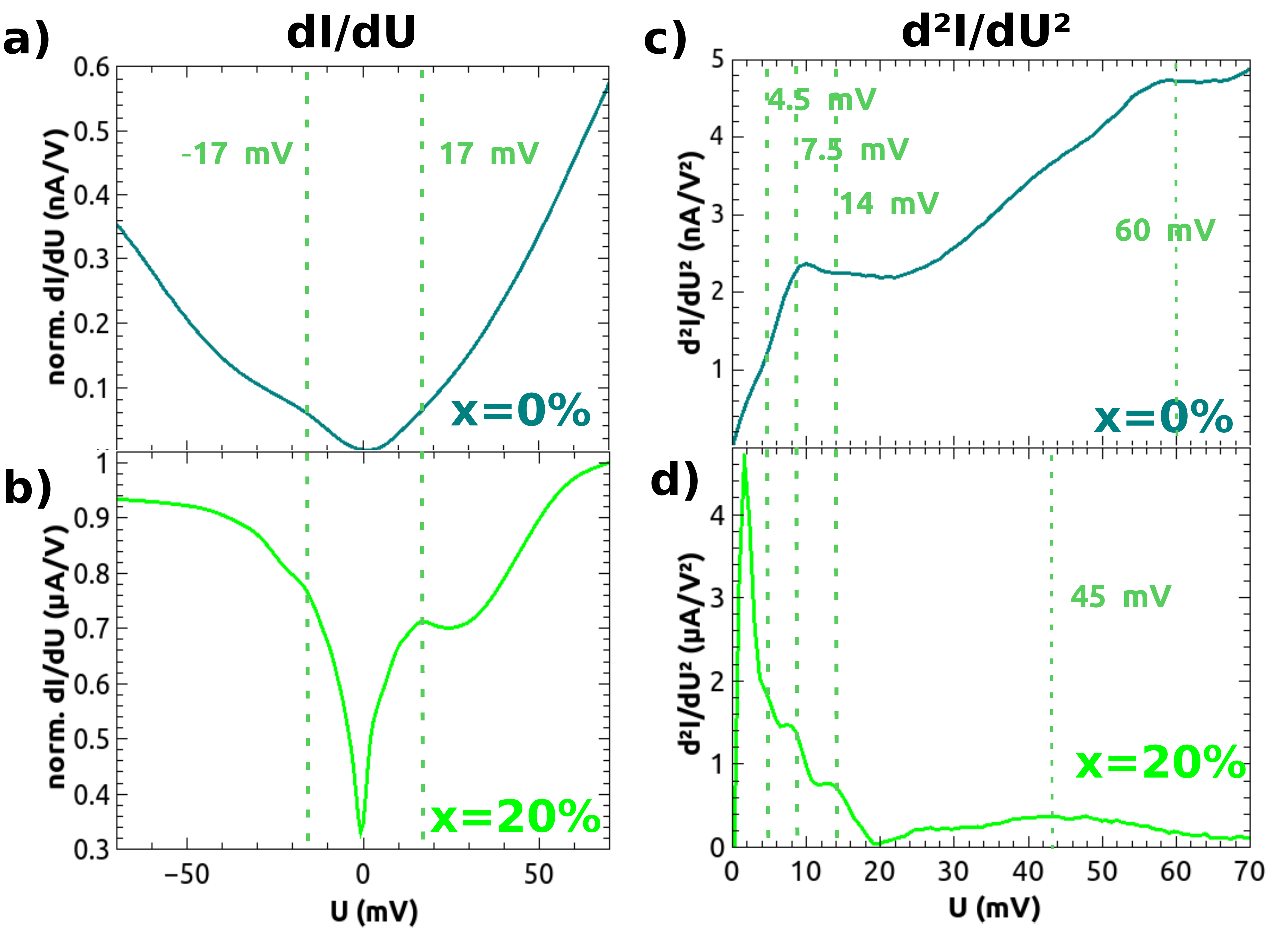}
\caption{(Color online) a) parent compound: 100 averaged $dI/dU$-spectra over an area of $20\times \unit[20]{nm^2}$ at $T=\unit[0.8]{K}$ ($\mathrm{U_m}$=\unit[2.9]{mV}) , b) magnetic compound with x=0.2: 90 averaged $dI/dU$-spectra over an area of $1.6 \times \unit[1.6]{nm^2}$ at $T=\unit[0.8]{K}$ ($\mathrm{U_m}=\unit[761]{\mu V}$), c) corresponding $d^2I/dU^2$-measurement of a), d) corresponding $d^2I/dU^2$-measurement of b)}
\label{inel2}
\end{figure}

Similar measurements were done for the magnetic compounds (x=0/0.2). Conductance measurements are shown in Fig. \ref{inel2} a) and b) whereas measurements of inelastic excitations are shown in Fig. \ref{inel2} c) and d). For the parent compound, the averaged dI/dU spectrum in Fig.\ref{inel2} a) shows a gap of $2\Delta =\unit[33.6]{meV}$ around the Fermi energy. Even though the gap is rather broad, we suggest that it represents a spin gap of $\Delta \approx \unit[17]{meV}$ due to nested electron bands causing a spin density wave \cite{Hsieh2008}. The same gap can be seen much clearer in the underdoped compound as shown in Fig. \ref{inel2}b).  

In Fig. \ref{inel2} c) and d) several inelastic excitations are visible either as clear peaks in the spectra or weak shoulders. In both samples we observe excitations at $\unit[4.5]{mV}$ possibly due to magnetic excitation of the spin density wave. At $\unit[7.5]{mV}$ both compounds show a feature which might be related to low energy (acoustic) phonons \cite{Kobayashi2011,Pintschovius2014}. Finally, both samples show a feature at $\unit[14]{mV}$, i.e. at similar energies as the excitations observed in the superconducting samples. We attribute these features to phonons \cite{Litvinchuk2008,Kobayashi2011}. Finally, we have observed broad features around $\unit[60]{mV}$ in the parent compound and around $\unit[45]{mV}$ in the underdoped compound. These energies are similar to the broad features found in the overdoped sample suggesting a magnetic origin, as well. Thus, we find back all of the excitations observed in the superconducting samples and in the magnetic samples linking the two phases.

In summary, we revealed a nodal superconducting gap for the optimally doped and overdoped compound of  SrFe$\mathrm{_2}$(As$\mathrm{_{1-x}}$P$\mathrm{_x}$)$\mathrm{_2}$ indicating a $s_\pm $-symmetry in this system. Furthermore, we determined the superconducting coherence length for the respective compounds of the order of a few nm.  While this is significantly smaller than the values found in conventional superconductors,  it is of the same order of magnitude compared to other pnictide superconductors. Due to the small coherence length, the local stoichiometry affects the superconducting properties \cite{Malozemoff2006}. Thus, for optimizing the superconductive properties, the doping needs to be homogeneous on the short length scales of the coherence length. The spectroscopic measurements indicate electron-phonon coupling in all four compounds. Besides phonons, we could identify magnetic excitations in the inelastic spectra giving further evidence for magnon-driven superconductivity.

\begin{acknowledgments}

The authors acknowledge funding by the DFG under project WU 349/12-1 and fruitful discussions with J\"org Schmalian and Patrik Hlobil.
 
\end{acknowledgments}

\end{document}